\documentclass[prd,aps,showpacs,nofootinbib,amsmath,amssymb]{revtex4}
\usepackage{amssymb,epsfig}

\newcommand{\Real}{\mathbb{R}}

\begin{document}

\title{Nonlinear instability of wormholes supported by exotic dust and a magnetic field}
\author{Olivier Sarbach and Thomas Zannias}
\affiliation{Instituto de F\'\i sica y Matem\'aticas,
Universidad Michoacana de San Nicol\'as de Hidalgo,\\
Edificio C-3, Ciudad Universitaria, 58040 Morelia, Michoac\'an, M\'exico.}

\begin{abstract}
Recently, spherically symmetric, static wormholes supported by exotic dust and a radial magnetic field have been derived and argued to be stable with respect to linear radial fluctuations. In this report we point out that these wormholes are unstable due to the formation of shell-crossing singularities when the nonlinearities of the theory are taken into account.
\end{abstract}

\date{\today}


\maketitle

\section{Introduction}

The most popular wormhole model \cite{mMkT88} is described by the spacetime $(M,{\bf g})$ where
\begin{equation}
M = \Real^2 \times S^2,\qquad
{\bf g} = -dt^2 + dx^2 + (x^2 + b^2)(d\vartheta^2 + \sin^2\vartheta\, d\varphi^2),
\label{Eq:EBMTWormhole}
\end{equation}
$(t,x)$ denoting Cartesian coordinates on $\Real^2$ and $(\vartheta,\varphi)$ the standard angular coordinates on $S^2$. Here, $b > 0$ is the areal radius of the throat which consists of the cylinder $x=0$. In order to satisfy Einstein's field equations, this solution needs to be supported by "exotic" matter, that is, matter fields which violate the null energy condition. Several such matter models have been proposed which admit the solution (\ref{Eq:EBMTWormhole}). For example, it has been shown \cite{hE73,kB73,cA02} that this solution may be supported by a ghost scalar field (a massless scalar field whose kinetic energy has a reversed sign). Also, the wormhole (\ref{Eq:EBMTWormhole}) may be sourced by an anisotropic perfect fluid with an appropriate equation of state \cite{jGfGnMtZ09}. More recently, it has been shown \cite{aSiNnK08} that this solution can also be supported by a combination of exotic dust (a pressureless fluid with a negative energy density) and a radial magnetic field.

Once a suitable matter model which admits the wormhole configuration (\ref{Eq:EBMTWormhole}) has been found, a relevant question is whether or not the solution is stable with respect to fluctuations of the metric and matter fields. Clearly, the stability properties depend on the specific matter model. In the case of black holes, for instance, the magnetically charged Reissner-Nordstr\"om black hole is linearly stable in Einstein-Maxwell theory \cite{vM74b}; however, it is unstable in Einstein-Yang-Mills-Higgs theory \cite{kLvNeW92} if the horizon is sufficiently small. In fact, a similar situation occurs for the Schwarzschild black hole when it is embedded in a spacetime with one extra compact dimension to form a black string. While the Schwarzschild solution is linearly stable in vacuum general relativity \cite{bKrW87} the corresponding black string is linearly unstable if the circumference of the extra dimension is large enough \cite{rGrL93}. Concerning the wormhole model (\ref{Eq:EBMTWormhole}), it has been shown to be unstable with respect to linear, radial perturbations \cite{jGfGoS09a} when coupled to the ghost scalar field. Numerical simulations in the nonlinear, spherically symmetric case indicate \cite{hSsH02,jGfGoS09b,aDnKdNiN09} that the wormhole either rapidly expands or collapses to a Schwarzschild black hole. Attempts have been made to stabilize the wormhole by adding an electric or magnetic charge \cite{jGfGoS09c}, resulting in charged generalizations of (\ref{Eq:EBMTWormhole}). Although the addition of charge reduces the time scale of the instability, linearly stable wormholes have not been found in the scalar field model.

In a recent article \cite{dNaDiNaS09} the linear stability of the wormhole (\ref{Eq:EBMTWormhole}) has been analyzed in the matter model put forward in  \cite{aSiNnK08} with the conclusion of stability. In this report, we point out that the wormhole (\ref{Eq:EBMTWormhole}) is unstable when nonlinear, radial fluctuations of the metric and the exotic dust are taken into account. The reason for the instability is that for a large class of initial data shell-crossing singularities form and therefore curvature invariants of the perturbed spacetime blow up in finite time.

\section{The model}

The model considered in \cite{aSiNnK08} consists of a spherically symmetric, self-gravitating dust cloud coupled to a radial magnetic field. In terms of a Lagrangian coordinate $x$ which parametrizes the dust shells and proper time $\tau$ measured by an observer comoving with such a shell, the key equation is a master equation\footnote{See the Appendix for a derivation.} for the areal radius $r(\tau,x)$ which, for each fixed shell $x$, has the form of a one-dimensional mechanical system,
\begin{equation}
\frac{1}{2} \dot{r}(\tau,x)^2 + V(r(\tau,x), x) = E(x),
\label{Eq:1DMechanical}
\end{equation}
with potential
\begin{equation}
V(r,x) = -\frac{\mu(x)}{r} + \frac{Q^2}{2 r^2}\; .
\end{equation}
Here, the energy function $E(x)$ is determined by the initial data $r_0(x) := r(0,x)$ and $v_0(x) := \dot{r}(0,x)$ for $r(\tau,x)$ at $\tau=0$, say, $Q$ is the magnetic charge and $\mu(x)$ is the mass function for the dust, which, in terms of the initial density $\rho_0(x)$ is
\begin{equation}
\mu(x) = \mu_\infty - 4\pi G\int\limits_x^\infty \rho_0(y) r_0(y)^2 r_0'(y) dy,
\label{Eq:DustMass}
\end{equation}
$\mu_\infty$ representing the mass at $x\to\infty$ and $G$ Newton's constant. Once the function $r(\tau,x)$ has been determined, the metric ${\bf g}$, the electromagnetic field tensor ${\bf F}$, the four-velocity ${\bf u}$ and the density $\rho$ are given by
\begin{eqnarray}
{\bf g} &=& -d\tau^2 + \frac{r'(\tau,x)^2}{1 + 2 E(x)}\; dx^2
 + r(\tau,x)^2(d\vartheta^2 + \sin^2\vartheta\, d\varphi^2),
\label{Eq:MetricSol}\\
{\bf F} &=& Q\, d\vartheta \wedge \sin\vartheta\, d\varphi,
\label{Eq:EMSol}\\
{\bf u} &=& \frac{\partial}{\partial\tau}\; , \qquad
\rho(\tau,x) = \rho_0(x)\left( \frac{r_0(x)}{r(\tau,x)} \right)^2\frac{r_0'(x)}{r'(\tau,x)}\; .
\label{Eq:FluidSol}
\end{eqnarray}
Here and in the following, a dot and a prime denote partial differentiation with respect to $\tau$ and $x$, respectively. Notice that there is a freedom in reparametrizing the dust shells, $x\mapsto f(x)$. In the Tolman-Bondi models (see, for instance, Ref. \cite{Papapetrou-Book}) this freedom is usually fixed by labeling each shell by its initial areal radius, $r_0(x) = x$. Since here we are interested in describing wormhole spacetimes with one throat, we set $r_0(x) = \sqrt{x^2 + b^2}$ instead, where $b > 0$ is the throat's initial areal radius.

Static solutions are obtained if each dust shell $x$ is in equilibrium with respect to the potential $V(r,x)$. This is possible if both $Q^2$ and $\mu(x)$ are strictly positive, in which case there is a potential well with global minimum at $r = Q^2/\mu(x)$ corresponding to the energy $E(x) = -Q^2/(2r^2)$. This, together with $r(x) = r_0(x) = \sqrt{x^2 + b^2}$ leads to the static wormhole solution (\ref{Eq:EBMTWormhole}) with $b=|Q|$, negative density
\begin{equation}
\rho_0(x) = -\frac{1}{4\pi G}\frac{Q^2}{(x^2 + Q^2)^2}
\label{Eq:rhoEquilibrium}
\end{equation}
and zero asymptotic mass $\mu_\infty=0$.

\section{Perturbation analysis}

Now let us consider a nonlinear perturbation of the static solution described in the previous section for a fixed value of the magnetic charge $Q$. Its Cauchy evolution is determined by the initial data for the functions $r_0(x)$, $v_0(x)$ and $\rho_0(x)$ and the value for the asymptotic mass $\mu_\infty$. Keeping the labeling of the dust shells such that $r_0(x) = \sqrt{x^2 + b^2}$, this amounts in perturbing $v_0(x)$, $\rho_0(x)$ and $\mu_\infty$ from their equilibrium values $v_0 = 0$, $\rho_0$ given by Eq. (\ref{Eq:rhoEquilibrium}) and $\mu_\infty=0$, respectively. The perturbations of $\rho_0$ and $\mu_\infty$ give rise to a new mass function $\mu(x)$ according to Eq. (\ref{Eq:DustMass}). For the following, we consider a rather large class of nonlinear perturbations which are "small" in the following sense:
\begin{enumerate}
\item[(i)] The perturbed mass function $\mu(x)$ is strictly positive for all $x\in\Real$.
\item[(ii)] The total energy $E(x)$ is strictly negative for all $x\in\Real$.
\end{enumerate}
Condition (i) implies that $V(\cdot,x)$ describes a potential well with negative global minimum at $r_{min}(x) = Q^2/\mu(x)$. Condition (ii) means that the trajectories of the system (\ref{Eq:1DMechanical}) are bounded; hence the areal radius of each dust shell 
undergoes periodic oscillations about $r_{min}(x)$. If $0 < r_1(x) < r_2(x)$ denote the zeroes of the function $r\mapsto E(x) - V(r,x)$ for each fixed $x$, the corresponding period is
\begin{equation}
T(x) = 2\int\limits_{r_1(x)}^{r_2(x)} \frac{dr}{\sqrt{2(E(x) - V(r,x))}} 
 = \frac{\pi}{\sqrt{2}} \frac{\mu(x)}{|E(x)|^{3/2}}\; .
\end{equation}
Since the static wormhole solution satisfies the conditions (i) and (ii) it is not difficult to 
construct initial data with these properties. For example, one could leave $\rho_0$ and $\mu_\infty$ unperturbed which also leaves the potential unperturbed and choose $|v_0|$ small enough such that $v_0(x)^2 < -2V(r_0(x),x) = Q^2/r_0(x)^2$ for all $x\in\Real$.

Since small perturbations fulfilling the conditions (i) and (ii) behave like a one-dimensional particle in a potential well one might expect the static wormhole ({\ref{Eq:EBMTWormhole}) to be stable under perturbations. However, a problem arises when different shells of dust touch or cross each other, in which case the density function $\rho(\tau,x)$ diverges. This occurs at points where the function $\nu(\tau,x) := r'(\tau,x)/r_0'(x)$ vanishes, see Eq. (\ref{Eq:FluidSol}). The time evolution of $\nu(\tau,x)$ is governed by the following equation which may be obtained by differentiating Eq.~(\ref{Eq:1DMechanical}) twice,
\begin{equation}
\ddot{\nu}(\tau,x) + \omega(\tau,x)^2\nu(\tau,x) 
 = -4\pi G\rho_0(x)\left( \frac{r_0(x)}{r(\tau,x)} \right)^2,
\end{equation}
where
\begin{displaymath}
\omega^2(\tau,x) = \frac{\partial^2 V}{\partial r^2}(r(\tau,x),x)
 = -\frac{2\mu(x)}{r(\tau,x)^3} + \frac{3Q^2}{r(\tau,x)^4}.
\end{displaymath}
Since $\nu(0,x) = 1$, $\dot{\nu}(0,x) = v_0'(x)/r_0'(x)$, shell-crossing is absent for small enough time provided that the initial function
\begin{displaymath}
\dot{\nu}(0,x) = \frac{dv_0}{dr}(x)
\end{displaymath}
is regular. In particular, this requires the initial velocity profile $v_0(x)$ to have a critical point at the throat. Otherwise, the time evolution of the wormhole is not defined since in this case
\begin{displaymath}
\dot{\rho}(0,x) = -\rho_0(x)\left[ 2\frac{v_0(x)}{r_0(x)} + \dot{\nu}(0,x) \right]
\end{displaymath}
diverges. The regularity of $\dot{\nu}(0,x)$ also implies the regularity of the radial part of the metric,
\begin{equation}
-d\tau^2 + \frac{r_0'(x)^2}{1 + 2E(x)} \nu(\tau,x)^2 dx^2
\end{equation}
whose Gaussian curvature is $\nu^{-1}\ddot{\nu}$, for small enough time. However, there is a coordinate singularity at $x=0$ unless $r_0'(x)^2/(1 + 2E(x))$ is finite and different from zero at $x=0$. This yields an additional restriction on the initial energy $E(x)$. In order to shed light on this condition, we assume for simplicity that $\rho_0$ and $\mu_\infty$ are unperturbed and set $v_0(x) = \alpha(x) |Q|/r_0(x)$ with some smooth function $\alpha(x)$ satisfying $\alpha(x)^2 < 1$ for all $x\in\Real$. Then, 
\begin{equation}
1 + 2E(x) = \frac{1}{r_0(x)^2}( x^2 + Q^2 \alpha(x)^2 ).
\label{Eq:PerturbedEnergy}
\end{equation}
Since $r_0'(x) = x/r_0(x)$, the coordinate singularity is avoided if $\alpha(x)$ is proportional to $x$ for small $|x|$. Together with the regularity of $\dot{\nu}(0,x)$ this means that the function $\alpha(x)$ should satisfy
\begin{equation}
\alpha(0) = 0,\qquad
\alpha'(0) = 0,\qquad
\alpha(x)^2 < 1, \quad x\in\Real.
\label{Eq:AlphaConditions}
\end{equation}

In the following, we assume that the above conditions on $\alpha(x)$ are satisfied in order for the coordinate $x$ to be globally defined on the real axis and in order to avoid shell-crossing singularities for arbitrary small times. To analyze the occurrence of shell-crossing singularities at later times, we differentiate the equation
\begin{displaymath}
r(\tau + n T(x),x) = r(\tau,x),\qquad 
n = 0,1,2,3,..., \quad \tau,x\in\Real,
\end{displaymath}
and obtain
\begin{equation}
\nu(\tau +  n T(x),x) = \nu(\tau,x) - n\dot{r}(\tau,x)\frac{T'(x)}{r_0'(x)}.
\end{equation}
The evaluation of this equation at a turning point $\tau = \tau_1(x)$, where $\dot{r}(\tau_1(x),x)=0$, yields
\begin{equation}
\nu(\tau_1(x) + nT(x),x) = \nu(\tau_1(x),x), \qquad
n = 0,1,2,3,..., \quad x\in\Real,
\label{Eq:rturnRperiodic}
\end{equation}
while for $\tau=0$ we obtain
\begin{equation}
\nu(nT(x),x) = 1 - n v_0(x)\frac{T'(x)}{r_0'(x)}, \qquad
n = 0,1,2,3,..., \quad x\in\Real.
\label{Eq:rRperiodic}
\end{equation}
Therefore, for each fixed shell $x$, the field $\nu(\tau,x)$ oscillates between the two values given in the right-hand side of Eqs. (\ref{Eq:rturnRperiodic}) and (\ref{Eq:rRperiodic}), respectively. If $v_0(x) T'(x)/r_0'(x) \neq 0$, this signifies that the oscillation's amplitude grows without bound for $\tau\to\infty$. Furthermore, since $\nu(0,x) = 1$, Eq. (\ref{Eq:rRperiodic}) also implies that for $v_0(x) T'(x)/r_0'(x) > 0$ a shell-crossing singularity must occur in finite proper time. It is a simple matter to construct initial data satisfying this condition everywhere away from the throat. For example, leaving $\rho_0$ and $\mu_\infty$ unperturbed  and setting $v_0(x) = \alpha(x) |Q|/r_0(x)$ with 
the function $\alpha(x)$ satisfying the restrictions (\ref{Eq:AlphaConditions}) gives
\begin{displaymath}
v_0(x)\frac{T'(x)}{r_0'(x)} = 4\pi\alpha(x)(1-\alpha(x)^2)^{-5/2}
\left[ 1 - \alpha(x)^2 + \frac{3}{2} r_0(x)^2\frac{\alpha(x)\alpha'(x)}{x} \right].
\end{displaymath}
If $\alpha_0(x)$ is any smooth function which satisfies the conditions (\ref{Eq:AlphaConditions}), is strictly positive for $x\neq 0$, and constant outside a compact interval, then the family of functions $\alpha_\lambda(x) = \lambda\alpha_0(x)$ with $\lambda > 0$ small enough is easily seen to satisfy $v_0(x) T'(x)/r_0'(x) > 0$ for all $x\neq 0$. We conclude that a large class of nonlinear perturbations leads to the formation of shell-crossing singularities. As a consequence, the wormhole is unstable.

We shall finish this report by making a few comments regarding the behavior of linear perturbations on the static wormhole background defined in the previous section. If we set $r(\tau,x)=r_{0}(x)+\delta r(\tau,x)$, differentiate Eq.~(\ref{Eq:1DMechanical}) with respect to $\tau$, we obtain, to linear order in $\delta r(\tau,x)$,
\begin{eqnarray}
 \frac{\partial^{2}\delta r(\tau,x)}{\partial\tau^{2}} + \omega_0(x)^{2}\delta r(\tau,x)=0,
 \qquad \omega_0(x)=\frac{|Q|}{r_{0}^{2}(x)},
\end{eqnarray}
and thus,
\begin{eqnarray}
\delta r(\tau,x)=\delta v_0(x)\frac{\sin(\omega_0(x)\tau)}{\omega_0(x)},
\end{eqnarray}
where we have assumed that $r_0$, $\rho_0$ and $\mu_\infty$ are unperturbed. For the following, we also assume that the linearized initial velocity $\delta v_0(x)$ is compactly supported away from the throat at $x=0$. The induced linear density perturbation $\delta\rho(x,\tau)$ is
\begin{eqnarray}
 \delta\rho(\tau,x)=\rho_0(x)\left[ -\frac{2\delta r(\tau,x)}{r_{0}(x)}-\frac{\delta r^{'}(\tau,x)}{r^{'}_{0}(x)} \right].
\end{eqnarray}
Since each shell oscillates with its own frequency $\omega_0(x)$, it  is seen that $\delta\rho(\tau,x)$ exhibits linear growth in time (see also the discussion in \cite{dNaDiNaS09}). In terms of the gauge-invariant combination
\begin{displaymath}
J(\tau,x):=\delta\rho(\tau,x) - \frac{\rho_0'(x)}{r_0'(x)}\delta r(\tau,x),
\end{displaymath}
this yields
\begin{equation}
J(\tau,x) = \frac{r_0(x)^3\rho_0(x)}{x} \left( \frac{\delta r(\tau,x)}{r_0(x)^2} \right)'
 = \frac{r_0(x)^3\rho_0(x)}{x} \left( \frac{\sin(\omega_0(x)\tau)}{\omega_0(x)}\frac{\delta v_0(x)}{r_0(x)^2} \right)'.
\end{equation}
Therefore, the linear growth in $\tau$ cannot be transformed away by an infinitesimal coordinate transformation. As we have shown above, in the nonlinear case, this growth leads to blowup in finite proper time.

\section{Conclusions}

The result of this report is that the wormhole (\ref{Eq:EBMTWormhole}), when supported by exotic dust and a radial magnetic field, is unstable due to the formation of shell-crossing singularities. Although shell-crossing singularities have been argued to be "mild" in the sense that it is often possible to construct a $C^1$-extension of the metric \cite{pSaL99}, nevertheless the density function and the Ricci scalar blow up. Even if one is willing to accept such singularities it is worth pointing out the following: according to the arguments presented below Eq. (\ref{Eq:rRperiodic}), the extended spacetime possesses the bizarre property that for large $\tau$, the norm of the vector field $\partial_x$, measuring the normal geodesic deviation between neighboring dust shells, undergoes wild oscillations.


\acknowledgments

We thank Dar\'{\i}o N\'u\~nez for interesting discussions. This work was supported in part by Grants No. CIC 4.7 and 4.19 to Universidad Michoacana, PROMEP UMICH-PTC-195 from SEP Mexico, and CONACyT 61173.

\appendix
\section*{Appendix: Derivation of the master equation}

The radial components of the Einstein equations for the metric
\begin{displaymath}
{\bf g} = -d\tau^2 + \frac{r'(\tau,x)^2}{\Gamma(\tau,x)^2}\; dx^2
 + r(\tau,x)^2(d\vartheta^2 + \sin^2\vartheta\, d\varphi^2),
\end{displaymath}
the four-velocity ${\bf u} = \frac{\partial}{\partial \tau}$ and the electromagnetic field ${\bf F} = Q\, d\vartheta \wedge \sin\vartheta\, d\varphi$ read
\begin{eqnarray}
G^\tau{}_\tau &=& \frac{2}{r} \frac{\dot{\Gamma}}{\Gamma} \dot{r}
 - \frac{2m'}{r^2 r'} = -\frac{Q^2}{r^4} - 8\pi G\rho,
\label{Eq:Gtt}\\
G^x{}_x &=& -\frac{2}{r}\ddot{r} - \frac{2m}{r^3} = -\frac{Q^2}{r^4},
\label{Eq:Gxx}\\
G^\tau{}_x &=& \frac{2}{r}\frac{\dot{\Gamma}}{\Gamma} r' = 0,
\label{Eq:Gtx}
\end{eqnarray}
where $G^{\alpha}{}_{\beta}$ are the components of the Einstein tensor and $m(\tau,R)$ is the Misner-Sharp mass function \cite{cMdS64} which is defined by
\begin{equation}
1 - \frac{2m}{r} = {\bf g}(dr,dr) = -\dot{r}^2 + \Gamma^2.
\label{Eq:Mass}
\end{equation}
These equations are solved as follows: first, Eq. (\ref{Eq:Gtx}) immediately implies that $\Gamma = \Gamma(x)$ is a function of $x$ only. Next, differentiating Eq. (\ref{Eq:Mass}) with respect to $\tau$ and comparing with Eq. (\ref{Eq:Gxx}) yields $\dot{m} = Q^2\dot{r}/(2r^2)$ which may be integrated to obtain
\begin{equation}
m(\tau,x) = \mu(x) - Q^2/(2r(\tau,x)),
\label{Eq:MSDMRelation}
\end{equation}
with $\mu(x)$ representing the dust mass function. Finally, using this in Eq. (\ref{Eq:Gtt}) gives $\mu' = 4\pi G\rho r^2 r'$. As a consequence, $\rho r^2 r'$ must be independent of $\tau$ which implies Eq. (\ref{Eq:FluidSol}), and Eq. (\ref{Eq:DustMass}) is obtained after integrating in $x$. The key equation (\ref{Eq:1DMechanical}) that governs the dynamics of the system is obtained from Eq. (\ref{Eq:Mass}) after setting $\Gamma(x)^2 = 1 + 2E(x)$ and taking into account the relation (\ref{Eq:MSDMRelation}) between the mass functions $m$ and $\mu$. The remaining components of the Einstein equations and the mass conservation law are automatically satisfied as a consequence of the previous equations and Bianchi's identities.

\bibliographystyle{unsrt}
\bibliography{refs}

\end{document}